# Superluminal spacetime boundary, time reflection and quantum light generation from relativistic plasma mirrors

Chenhao Pan[1,2], Xinbing Song[3], Yang Cao[4], Li Xiong[5], Xiaofei Lan[5], Shaoyi Wang[6*], Yuxin Leng[1,2*], Yiming Pan[1*]

[1]*School of Physical Science and Technology and Center for Transformative Science, ShanghaiTech University, Shanghai 201210, China*
[2]*State Key Laboratory of High Field Laser Physics and Chinese Academy of Sciences Center for Excellence in Ultra-Intense Laser Science, Shanghai Institute of Optics and Fine Mechanics, Chinese Academy of Sciences, Shanghai 201800, China*
[3]*School of Physics, Beijing Institute of Technology, Beijing 100081, China*
[4]*Physics Department, Technion, Haifa, 320000, Israel*
[5]*School of Physics and Astronomy, China West Normal University, Nanchong 637002, China*
[6]*Science and Technology on Plasma Physics Laboratory, Research Center of Laser Fusion, China Academy of Engineering Physics, Mianyang 621900, China*

**Abstract**

A plasma mirror is an optical device for high-power, ultrashort-wavelength electromagnetic fields, utilizing a sheet of relativistic oscillating electrons to generate and manipulate light. In this work, we propose that the spatiotemporally varying plasma oscillation, induced by an ultra-high-intensity laser beam, functions as a "spacetime mirror" with significant potential for exploring quantum light. We find that the spacetime mirror exhibits several exotic features: (i) a superluminal spacetime boundary, (ii) time reflection and refraction, and (iii) quantum light sources with pair generation. Our theoretical and simulation results are in excellent agreement, and experimental verification is underway. Our work demonstrates the interplay with emerging fields such as time varying media, suggesting the plasma mirror as an ideal platform to study strong-field quantum optics at extremes.

A mirror is an optical device that reflects light waves toward the observer, forming an image of the object or source. The reflection occurs due to a sharp change of the refractive index in space. Typically, a plane mirror consists of a substrate made of metal or dielectric material, creating a space boundary that reflects incident electromagnetic waves. Beyond its passive function, a flying mirror or an oscillating mirror can actively control light waves. For instance, a flying mirror can Doppler-shift the frequency of the reflected waves [1], potentially enabling the creation of bright X-ray source in the future [2]. Meanwhile, an oscillating mirror uses relativistic dense plasma to generate extreme ultraviolet or x-ray harmonic radiations and enhance the temporal contrast of high-intensity laser beams [3–9]. Such a flying or oscillating mirror is commonly referred to as a plasma mirror, and its experimental investigation is crucial for relativistic plasma optics [10–12].

When an intense femtosecond laser pulse ($> 10^{18} Wcm^{-2}$) strikes a solid target, it ionizes the target, forming a dense plasma sheet at the surface. This plasma reflects the incoming beam, acting like a mirror. Since the 1980s, various ideas have been proposed to utilize the plasma mirror to explore strong-field and attosecond physics in ultrashort-wavelength regimes [3–9,13]. Due to its ability to withstand extremely high intensities, the plasma mirror allows for controlling ultra-high intensity laser beams. Recent advancements have opened up new directions, offering applications such as improving laser intensity contrast [14], and creating plasma analogues of optical elements like gratings [15], modulators [16]. Additional applications involving ionized electrons include electron acceleration [17–19]. These advancements have been reviewed [12,20]. Returning to our primary concern, we pose a question: Can the oscillating plasma sheet generate or manipulate quantum light? If so, can we extend plasma optics into the quantum regime?

At first glance, the question might seem nonsensical since laser-irradiated plasma is highly dense and extremely hot, composed of relativistic electrons oscillating collectively and colliding chaotically like classical particles. The laser-plasma interaction is strong-field and non-perturbative, suggesting that the quantum nature of relativistic plasma cannot persist under such extreme circumstances. To date, there have been few theoretical investigations into X-ray quantum optics [21] and even fewer experimental studies [22]. However, upon closer examination of the plasma sheet dynamics, its mechanism can be understood as that of an active mirror [11,12] . Previous studies have proposed that an accelerated mirror create quantum light emission as a squeezing effect [23–25], closely linked to a bunch of ideas such as parametric resonance [26] and dynamical Casimir effect [27,28]. Also, a recent work by Chen and Mourou [21] follows this inquiry, utilizing a plasma mirror to investigate the analogous dynamics of black hole evaporation and Hawking radiation.

As an oscillating mirror, we recognize a plasma mirror as a time-varying medium. Time-varying media involve sudden changes or periodic modulations of the dielectric constant to manipulate wave propagation, which have recently gained widespread attentions [29–31]. These temporal modulations enable applications such as intensity amplification [30], temporal switching [32] and aiming [33], as well as time reflection [34] and frequency conversion. The main challenge in experimenting with such media is their slow modulation speed for manipulating waves with short wavelength. Identifying efficient materials, such as epsilon-near-zero (ENZ) medium, and high-index dielectrics [35–38], remains a challenge. For instance, a recent study showed that an ENZ-based time-varying mirror extend the reflected wave frequency up to 31 THz [39]. Instead, most demonstrations have been in water waves [34] and microwaves [40]. The genuine mechanism for temporal modulation of optical waves remains elusive. Addressing this challenge, we realize that plasma mirrors offer a promising platform for testing time-varying media at an optical regime.

To this point, we propose that plasma mirror can serve as a spacetime mirror, exhibiting a plasma-vacuum interface that varies spatiotemporally in the optical and even X-ray regimes. This plasma-based spacetime mirror possesses both subluminal and superluminal boundaries. While plasma mirrors are widely discussed for high harmonic generations, our focus here is on quantum light source and time reflection. Several key quantum features of the spacetime mirror are demonstrated, including entangled photon generation via the squeezing effect on a solid target, and coincident measurement of quantum light generation and high harmonics generation. Our work paves a paradigmatic path for engaging ultra-high-intensity plasma physics, quantum optics, and time-varying media, potentially leading to a new field of quantum plasma optics and the exploration of pair generation and other quantum phenomena under extreme conditions.

**Modeling spacetime mirror**

Initially, we conceptualize the plasma mirror as a Relativistic Oscillating Mirror (ROM) and construct a spacetime mirror based on this framework. The ROM model, introduced in the [8,9], elucidates the interaction between high-intensity lasers and plasma. This interaction entails the laser inducing a sharply defined vacuum-plasma interface, which it subsequently drives to oscillate at relativistic speeds. Within the realm of strong-field plasma physics, free electrons in the plasma mirror are described by classical electrodynamics. Our proposed spacetime mirror similarly adheres to this classical paradigm but would eventually demonstrate its quantum nature.

Unlike the interface within the ROM model, we emphasize that the oscillation velocity of the vacuum-plasma interface can exceed the speed of light. In this regime, the interface functions as a time boundary rather than a space boundary. This time boundary can induce non-classical modulations on the electromagnetic wave [41], despite the classical plasma nature of the boundary itself. In general, we assert that the spacetime mirror can exhibit both space and time reflections.

A linearly polarized high-intensity laser is employed to drive a high-density plasma, thereby constructing the ROM model. The laser field is represented as $\boldsymbol{A}(t) = A_0(t)\sin(\omega_L t - k_L x + \phi_c)$, where $A_0(t) = A_0 \exp\left(-\frac{t^2}{(2t_p)^2}\right)$ gives the temporal envelope of the laser intensity, with $2t_p$ being the pulse duration, $A_0$ the peak intensity, $\omega_L$ the angular frequency, $k_L$ the wave vector, and $\phi_c$ the carrier envelope phase of the laser. The plasma sheet forms a sharp interface with the vacuum, described by $n(x) = n_0 \Theta(x - \lambda_s(t))$ with $\Theta(x)$ being the Heaviside function. The expression $\lambda_s(t)$ denotes the motion of the interface. $n_0 = n_e/n_c$ is the normalized plasma density, $n_e$ is the laser-induced electron density and $n_c = m_e \varepsilon_0 \omega_L^2 / e^2$ is the plasma critical density with $m_e, e$ are the electron mass and charge, and $\varepsilon_0$ is the vacuum permittivity. We note that the plasma density $n_e$ is approximated to be constant within the ROM model. The interface follows the equation:

$$\frac{d}{dt}(\dot{\lambda}_s(t)\gamma) = -n_0 \lambda_s(t)\Theta(\lambda_s(t)) + \frac{a}{\gamma}\frac{\partial}{\partial x}a \qquad (1)$$

where $\gamma = \sqrt{1 + |a_o(t)|^2}/\sqrt{1 - \dot{\lambda}_s(t)}$ is the Lorentz factor. A detailed derivation of Eq. 1 can be found in the SM file. For simplicity, the normalized laser field strength is denoted as $a(t) = eA_0(t)/m_e\omega_L c$. Initially, in the absence of external driving (i.e., $a = 0$), the interface function is stable $\lambda_s(t) = 0$ and the oscillation velocity of the interface is zero $\dot{\lambda}_s(t) = 0$.

The oscillatory behavior of the vacuum-plasma interface arises from the interplay between the ponderomotive force and Coulomb force of plasma. Upon the laser interacting with the interface, the Coulomb force (the first term of LHS) and the ponderomotive force of strong laser (the second term of LHS) exert opposite influences at the first half of optical cycle. The Coulomb force cannot immediately accumulate to balance the ponderomotive force results in unidirectional acceleration of the target surface before the Coulomb force accumulates to balance the acceleration. At the saturation case $\frac{d}{dt}(\dot{\lambda}_s(t)\gamma) = 0$, signifying no boundary oscillation, the equation $\frac{a}{\gamma}\frac{\partial}{\partial x}a = n_0\lambda_s(t)\Theta(\lambda_s(t))$ describes the balance between the ponderomotive force and the Coulomb repulsion, marking the maximum velocity by the boundary. During the subsequent half optical cycle, the ponderomotive force inverts its direction while the direction of Coulomb force remains, inducing deceleration of the interface followed by acceleration in the opposite direction. Post the interface surpassing its initial position, the Coulomb force reverses to balance the ponderomotive.

Despite the analytical solutions of this nonlinear equation, the boundary motion can be approximated through Fourier series as the ponderomotive force dominates as the sole external source.

$$\lambda_s(t) \simeq \sum_l A_l(t) \sin(2l\omega_L t) \tag{2}$$

Here, $l$ denotes the order of harmonic oscillation, and $A_l(t)$ represents the amplitude of the $l$ th order oscillation. The temporal variations of the boundary function correspond to periodic alterations in the refractive index within its oscillation region, transitioning from vacuum ($n_0 = \sqrt{\varepsilon_0} = 1$) to plasma ($n_p = \sqrt{1 - n_0/n_c}$), where $n_p$ can be derived from Drude model. For $n_0 > n_c$, light undergoes near-total reflection, analogous to mirror, forming the basis for the ROM effect [42]. Given a specific laser pulse and plasma density, the reflected wave from the interface is approximated as $E_{HHG} = E_0(t) \sin\left(\omega_L t + \frac{\omega_L \lambda_s(t)}{c}\right) \simeq \sum_{l=0} J_l(k_L A_s) \sin((2l+1)\omega_L t)$, where $J_l$ denotes the Bessel function of the first kind, and $E_0(t)$ is the initial pulsed field envelope. The emergence of new frequencies from plasma mirror is attributed to high-harmonic generation (HHG), facilitated by the relativistic speeds of the oscillating interface, given by

$$\dot{\lambda}_s(t) \simeq \sum_l 2l\omega_L A_{s,l}(t) \cos(2l\omega_L t) \tag{3}$$

where $\lambda_L = 2\pi c/\omega_L$ is the laser wavelength, and the amplitudes $A_{s,l}$ is estimated around $(1/10)\lambda_L$ with numerical and PIC simulations. This estimation indicates predominately relativistic speeds for the oscillating mirror. However, the superposition of multiple harmonic components on the vacuum-plasma interface can induce transient superluminal behavior, which is at special time, $\dot{\lambda}_{s,max} = \sum_l 2l\omega_L |A_{s,l}(t)| > c$. It is essential to clarify that while the boundary oscillates at superluminal velocities [43], this does not imply superluminal motion of electrons. The oscillating plasma-vacuum interface is consisted of a bunch of ionized collective electrons. These electrons that moves in a luminal velocity can allow the beam trajectories superluminal, analogous to an optical caustic. The vacuum-plasma interface is viewed as a varying spacetime mirror for controlling light beams, characterized by the mirror's oscillation velocity that can exceed or fall below the speed of light. In this sense, we emphasize that the ROM functions a spacetime mirror, capable of inducing both spatial and temporal reflections (Figure 1 (b) and (c)), thereby establishing a platform for controlling ultra-short-wavelength light propagation. In the superluminal regime, the interaction between light and the vacuum-plasma interface can results in time reflection (and time refraction), enabling the non-classical light generation (such as pair generation) even though the description of the superluminal interface adheres to classical frameworks.

**Generating quantum light from spacetime mirror**

The interaction between laser and plasma is revisited from the perspective of a spacetime mirror, exhibiting two distinct regimes: the subluminal regime, characterized by subluminal oscillation of the vacuum-plasma interface, and the superluminal regime, where the interface oscillates faster than the speed of light. In the superluminal regime, the spacetime mirror functions as a time-varying mirror, with the capacity to generate quantum light [30,41]. Unfortunately, the detection of quantum emission is blurred by the reflection of the incoming strong-field light. Isolating the time-reflected non-classical wave requires filtering out both the incident field and the reflected high harmonics. This filtering process, however, would also eliminate the time-reflected radiation due to the overlapping domains in frequency and space along the refection direction. Notably, the plasma sheet itself naturally filters the incident and reflected fields, preventing them from penetrating the target. Moreover, the plasma interface can exert time-varying modulation on the electromagnetic field within the un-ionized bulk of the target, potentially generating quantum light unaffected by the plasma screening.

To effectively detect quantum emission from the spacetime mirror, we designed an experiment that concurrently measures high harmonic generation (HHG) in front of a solid target and quantum light behind it. A 200TW mid-infrared laser ($\lambda_L = 800nm$) is employed to produce an ultra-short, ultra-intense laser pulse ($30fs$ duration and peak intensity of $10^{20}W/cm^2$). This pulse is focused to a 4μm spot using an off-axis parabolic (OAP) mirror and is directed at the target (Figure 2). The ultra-high intensity is crucial for generating high-density plasma on the target surface, while the ultra-short duration mitigates thermal accumulation. To filter the incident field, we use a solid target with a 100-nanometer thick metallic coating, which increases the plasma density ($n_0 \gg n_c$). The elevated plasma density effectively prevents the penetration of both incident and scattered waves.

In order to observe quantum emission effectively, we designed a sandwich structure for further amplifying the radiation. The main structure consists of a transparent dielectric material (e.g., $SiO_2$), with its rear surface coated with a high-reflective film, forming a resonant cavity. The cavity length is set at 800nm, corresponding to the optical wavelength $\lambda_L$ of the incident laser. The high-reflective film enables the partial transmission of quantum emissions created within the cavity. As illustrated in Figure 2 (b) inset left above, a sandwich cavity formed by the oscillating plasma sheet at the front of the dielectric target and the high-reflective film at its rear. The requirement of this design is two-fold: (i) preventing the penetration of incoming light to ensure the initial cavity modes at vacuum, and (ii) maintaining the dielectric constant of the sandwich structure uniform and stable for effective quantum light generation.

In examining the vacuum state of the electromagnetic field, we employed a quantized framework [44] to describe the field dynamics inside a cavity with oscillating boundaries. The detailed derivation can be found in Section 2 of the SM file. We note that the laser-induced plasma sheet, acting as an oscillating boundary, facilitated the quantization of the electromagnetic field as represented by the following Hamiltonian:

$$H_{EM} = \hbar \sum_n \omega_n(t) a_n^\dagger a_n + i \sum_k \xi_n(t)\left(a_n a_n - a_n^\dagger a_n^\dagger\right)$$
$$+ \frac{i}{2} \sum_{\substack{m,n \\ m \neq m}} \mu_{m,n}(t)\left(a_n^\dagger a_m^\dagger - a_m a_n + a_m^\dagger a_n - a_m a_n^\dagger\right) \quad (4)$$

The operators $a_n^\dagger(a_n)$ are the creation (annihilation) operators for cavity modes with wave vector $\boldsymbol{k_n}$. The time-dependent coupling factors $\xi_n(t)$ and $\mu_{m,n}(t)$ characterize interactions between modes that are associated with the oscillating interface ($\lambda_s(t)$), of which the explicit expressions can be found in the SM file. The cavity mode's frequency is $\omega_n(t) = n\pi c/L(t)$, where $n$ is an integer, $L(t) = L_0 - \lambda_s(t)$ represents the varying cavity length, $L_0$ is the position of the high-reflective film, and $\lambda_s(t)$ accounts for plasma sheet oscillation amplitude around the static front of the target, typically approximating $\lambda_L/10$. Note that without oscillating boundary when $L(t) \to L_0$, the above Hamiltonian simplifies to that of the EM field in a resonant cavity with $\xi_n(t) \to 0$ and $\mu_{m,n}(t) \to 0$.

To gain insight into the behavior of the EM field within an oscillating cavity, we proceed to investigate the evolution of the cavity field. We represent the Hamiltonian in the interacting picture via performing an unitary operator $U_0(t) = \exp{-i \int dt H_0(t) t}$, with $H_0(t) = \hbar \sum_n \omega_n(t) a_n^\dagger a_n$, and $\omega_n \approx n\omega_L$ since $\lambda_s \ll 1$. Thus, the Hamiltonian is rewritten by:

$$H_{tot} = H_1^I + H_2^I + H_3^I \tag{5}$$

with each terms expressed as:

$$H_1^I = i \sum_n \xi_n(t) \left( a_n^{\dagger 2} e^{i(2\omega_n)t} - a_n^2 e^{-i(2\omega_n)t} \right) \tag{6a}$$

$$H_2^I = \frac{i}{2} \sum_{\substack{m,n \\ m \neq n}} \mu_{m,n}(t) \left( a_n^\dagger a_m^\dagger e^{i(\omega_n+\omega_m)t} - a_n a_m e^{-i(\omega_n+\omega_m)t} \right) \tag{6b}$$

$$H_3^I = \frac{i}{2} \sum_{\substack{m,n \\ m \neq n}} \mu_{m,m}(t) \left( a_n^\dagger a_m e^{i|\omega_n-\omega_m|t} - a_n a_m^\dagger e^{-i|\omega_n-\omega_m|t} \right) \tag{6c}$$

Correspondingly, the coefficients can be explicitly derived from the interface (Eq. 2)

$$\xi_n(t) \approx \sum_l \sum_n \left( \frac{l\omega_L A_{s,l}}{4L_0} \right) \exp{i(2l\omega_L)t} \tag{7a}$$

$$\mu_{m,n}(t) = \sum_l \sum_{\substack{m,n \\ m \neq n}} (-1)^{(m+n)} \frac{mn}{m^2-n^2} \left(\frac{m}{n}\right)^{\frac{1}{2}} \left( \frac{l\omega_L A_{s,l}}{2L_0} \right) \exp{i(2l\omega_L)t} \tag{7b}$$

We leave the detailed derivations in the SM file (see Section 2). Using the rotating-wave approximation, we can further obtain the explicit expressions of the full Hamiltonian $H_{tot} = H_{1,RWA}^I + H_{2,RWA}^I + H_{3,RWA}^I$:

$$H_{1,RWA}^I \approx i \sum_n \left( \frac{n\omega_L A_{s,2n}}{4L_0} \right) \left( a_n^{\dagger 2} - a_n^2 \right) \tag{8a}$$

$$H_{3,RWA}^I = \frac{i}{2} \sum_{\substack{m,n \\ m \neq n}} (-1)^{(m+n)} \frac{mn}{j^2-k^2} \left(\frac{n}{m}\right)^{\frac{1}{2}} \left( \frac{(m+n)\omega_L A_{s,m+n}}{2L_0} \right) \left( a_n^\dagger a_m^\dagger - a_n a_m \right) \tag{8b}$$

$$H_{3,RWA}^{I} = \frac{i}{2} \sum_{\substack{m,n \\ m \neq n}} (-1)^{(m+n)} \frac{mn}{j^2 - k^2} \left(\frac{n}{m}\right)^{\frac{1}{2}} \left(\frac{(m-n)\omega_L A_{s,m-n}}{2L_0}\right) (a_n^\dagger a_m - a_n a_m^\dagger) \quad (8c)$$

These approximated Hamiltonian reveals two non-adiabatic processes. The first process, in which no photons are generated, is described by the scattering term $a_n^\dagger a_m$ in the Hamiltonian $H_{3,RWA}^{I}$. The second process involves pair generation, described by the terms $a_n^\dagger a_m^\dagger$ and $a_n^{\dagger 2}$ in the Hamiltonians $H_{1,RWA}^{I}$ and $H_{2,RWA}^{I}$, wherein a pair of photons are created. The efficiency of these paired photon generation crucially depends on the magnitudes of the laser-induced coefficients $\xi_n(t)$ and $\mu_{m,n}(t)$.

To ascertain which non-adiabatic process dominate within our proposed setup, we need to estimate the magnitudes of $\xi_n(t)$ and $\mu_{m,n}(t)$ from the parameters of the plasma sheet (see Section 2 in the SM file). Our analysis revealed that $\xi_n(t) \gg \mu_{m,n}(t)$ for the first few harmonic orders. Given that the higher-order amplitudes $A_{s,l}$ are inherently weak, we solely consider the dominant single−mode photon pair generation process in Eq. (8a). Consequently, the expectation value of the number operator of the $k^{th}$ harmonic mode can be approximated as:

$$\langle \mathcal{N}_k \rangle = \langle 0 | \exp(iH_{1,RWA}t) \, a_k^+ a_k \exp(-iH_{1,RWA}t) | 0 \rangle \quad (9)$$

where the $n^{th}$ harmonic is selected by the condition $n\omega_L = \omega_k$, and the total photon number generated from the vacuum $\mathcal{N}_n(t) = \sinh^2[\xi_n(t)t]$, exhibits exponential growth over time, as shown in Figure 3 (b) inset. This phenomenon is known as a squeezing effect [41,44], and consistent with the dynamic Casimir effect that have previously observed in microwave system with superconductor circuits [45]. Notably, our proposed setup allows for observing quantum photon emission in the optical or even extreme ultraviolet range, given that $\xi_n(t)$ incorporates the general multiple high harmonic frequencies associated with extremely shot wavelengths. Besides, the number of generated photons is determined by the interaction time $t$, the oscillating amplitude $\lambda_s(t)$ and its velocity $\dot{\lambda}_s(t)$. The interaction time is governed by the pulse duration, whereas the oscillating amplitude and velocity are related to the incident laser field strength $a_0$ and the induced plasma density $n_0$, as outlined in Eq. (1).

**Numerical simulations**

To demonstrate these predicated generation of non-classical photons and their dependence on these parameters, we conduct numerical simulations, with the results presented in Figure 3. Initially, we investigate the relation between the photon number and the laser field strength $a_0$ and plasma density $n_0$, while fixing the interaction duration at laser half-pulse width $t_p = 10T_L$. Figure 3(a) shows that generally, increasing laser intensity and decreasing plasma density led to greater photon pair generation. Notably, at very low plasma density, significant fluctuation emerges. These fluctuations arise from insufficient Coulomb interaction in the low-density region to counterbalance the ponderomotive force, resulting in deformation of the vacuum-plasma interface, which may result in the transmission of the laser field. Such deformation hinders the high-frequency oscillation of the interface, indicating a maximal harmonic cut-off. Conversely, a further increase in plasma density does not invariably increase photon number yield; a substantial decrease in photon number occurs at extremely high plasma densities. This feature is attributed to the strong Coulomb forces in high-density plasma, which suppress the laser-induced oscillation of the interface. As shown in Figure 3(a), we select parameters from the regions of ascending or plateau to optimize the experimental conditions.

Figure 3(b) demonstrate an exponential increase in photon production by extending the interaction time. The parameters $a_0 = 10$ and $n_0 = 100n_c$ are selected for the simulation. As the laser pulse duration extends from 4 to 12 optical cycles, the number of generated photons rises exponentially, from 0.1 to 4 photons. The inset provides a detailed visualization of this exponential growth.

Quantum emission is produced through the vacuum squeezing mechanism of the vacuum-plasma interface, whereas high-harmonic generation (HHG) results from its reflection. To explore the correlation between HHG and quantum emission, we select a specific high-frequency harmonic modes originating from the same oscillation pattern of the interface. Under the previous parameters, we extract the spatial reflections of the interface and perform a Fourier transform. Figure 4(b) reveals that both spectra exhibit a similar power law distribution, with HHG showing only odd harmonics and the interface involving even harmonics. From a classical perspective, this occurs because the HHG frequencies result from the interplay between the fundamental frequency and the induced oscillation frequency of the interface. This connection allows us to infer the dynamics of the plasma-vacuum interface through HHG spectrum analysis.

Quantum theory further predicts that under resonance conditions, the frequency of photons generated by the spacetime mirror is typically half its oscillation frequency, with photon number production proportional to the interface oscillation amplitude. Figure 4(d) demonstrate the Wigner function distribution corresponding to second and fourth-order interface oscillation on vacuum squeezing. Notably, higher frequency oscillations exhibit reduced amplitudes, thereby diminishing the effectiveness of vacuum squeezing and paired photon generation.

To characterize the quantum nature of this photon pairing, we analyzed the quantum statistics and entanglement properties of the emitted photons. Focusing on the second harmonic oscillations of the interface, we observe that the statistical distribution. Figure 4(a) exhibits a non-classical characteristic of the sub-Poissonian distribution. Given that vacuum squeezing produces photon pairs, symmetrical regions behind the target can simultaneously detect two correlated photons within a time window corresponding to the interaction time Figure 2 (inset above left), with the measured directional angle in accordance with momentum conservation.

Therefore, these photon pairs are expected to exhibit strong entanglement, which is readily detectable using a Hong-Ou-Mandel (HOM) interferometer, a typical measurement approach in quantum optics [45,46]. As depicted in Figure 2 right panel, two single-photon detectors are symmetrically placed behind the target to collect these paired photon emissions. The HOM interferometer path is employed to analyze the non-classical photon statistics. For entangled photons, adjusting the delay between pulses results in destructive interference at the beam splitter, yielding a zero-coincidence count at zero delay. Conversely, a maximum coincidence count of one is observed for delays exceeding the incoming laser pulse duration. The numerical results Figure 4 (c) show a zero-coincidence count, thereby confirming the photon entanglement produced through the vacuum squeezing of the proposed spacetime mirror.

**Further discussions**

In this section, we provide several discussions of the experimental requirements for realizing a plasmonic spacetime mirror. The realization of the plasma mirror can be achieved using most petawatt (PW) class ultra-short, ultra-intense laser systems [47,48]. The setup required for generating and detecting high-harmonic generation (HHG) in the ROM regime is well-established [49–51]. By comparison, quantum light measurement has been in use for a longer history and are more widely implemented [52–55]. However, the experimental challenge arises in the separation of both signals from the strong-field and quantum light measurements. This necessitates the selection of an appropriate target that can enhance and differentiate the strong-field and quantum light signals. As shown in Figure 2 (inset, bottom right), we expect that relative low laser intensity and high plasma density can help achieving optimal measurement conditions.

Despite the promising potential of the spacetime mirror, significant challenges persist, particularly in the generation of entangled photon pairs with high photon counts. Our theoretical framework suggests that extending the interaction time between the laser and plasma, along with using a medium with an optimal plasma density, can be an effective strategy. However, the strategy poses considerable challenges for current laser technology for current laser technology, as producing high-repetition-rate, high-contrast, stable ultra-short and ultra-intense pulses on the scale of hundreds of femtoseconds remains difficult [56]. This difficulty hampers the stable and precise control of laser-plasma interactions, leaving a gap in achieving reliable, high-repetition-rate operation for the spacetime mirror as a quantum source. Ideally, it is feasible for the spacetime plasma mirror to directly generate entangled photons in the XUV range through vacuum squeezing. Nevertheless, our numerical results suggest that, according to the traditional HHG power law, the vacuum squeezing capability of the spacetime mirror in the XUV range is relatively weak. Recently, an anomalous high-order enhanced HHG spectra have been observed in ROM experiments [57], indicating that the ROM mechanism has the potential for high harmonics control and the XUV/X-ray, even to the gamma ray pair generation.

**Conclusions**

In brief, we developed a superluminal spacetime mirror model based on the interaction between ultra-intense, ultra-short lasers and plasma. Through this classical model, we find that the induced vacuum-plasma interface exhibits superluminal oscillations under certain conditions, making the generalized Snell's law inapplicable and replacing it with time reflection and time refraction. We propose that the spacetime mirror can modulate the incident laser and generate high-harmonic generation (HHG). Meanwhile, at the superluminal region, it has the potential to serve as a high-quality quantum light source. We demonstrated how the spacetime mirror can squeeze the vacuum to produce entangled photons with short wavelengths, which has numerous applications in quantum sensing, and ultrafast quantum imaging.

Our theoretical framework and proposed experimental setup are designed to bridge the gap between strong-field laser plasma physics --- traditionally described by classical theories --- and quantum optics, through the use of the superluminal spacetime mirror. While this endeavor is challenging, it holds significant promise. Historically, the interaction between intense lasers and plasma has been recognized as an excellent source of compact and bright XUV radiation [49], effectively complementing the limitations inherent in XUV quantum sources, such as low photon yields [58,59] and restricted spectral ranges [60]. Nonetheless, plasma has typically been regarded as a hot, noisy electron fluid, generally deemed unsuitable for quantum treatment or for serving as a high signal-to-noise ratio quantum light source. Consequently, we aspire that the superluminal spacetime mirror will provide a promising connection between laser plasma physics and quantum optics, which distinguishes itself from the relativistic oscillating mirror and time-varying media, offering a unique platform capable of simultaneously supporting ultrafast strong-field exploration and non-classical quantum control at extremes.

**Acknowledgements:**

We would like to thank Prof. Jingwei Wang, Dr. Bin Zhang for insightful discussions. We appreciate for Dr. Qiaofei Pan's assistance of plotting figures. Y.P. is supported by the National Natural Science Foundation of China (Grant No. 2023X0201-417).

The authors declare no competing financial interests.

Correspondence and requests for materials should be addressed to Y.P. (yiming.pan@shanghaitech.edu.cn), Y.L.(lengyuxin@mail.siom.ac.cn) and S.W. (cslr1113@163.com).

**Figures:**

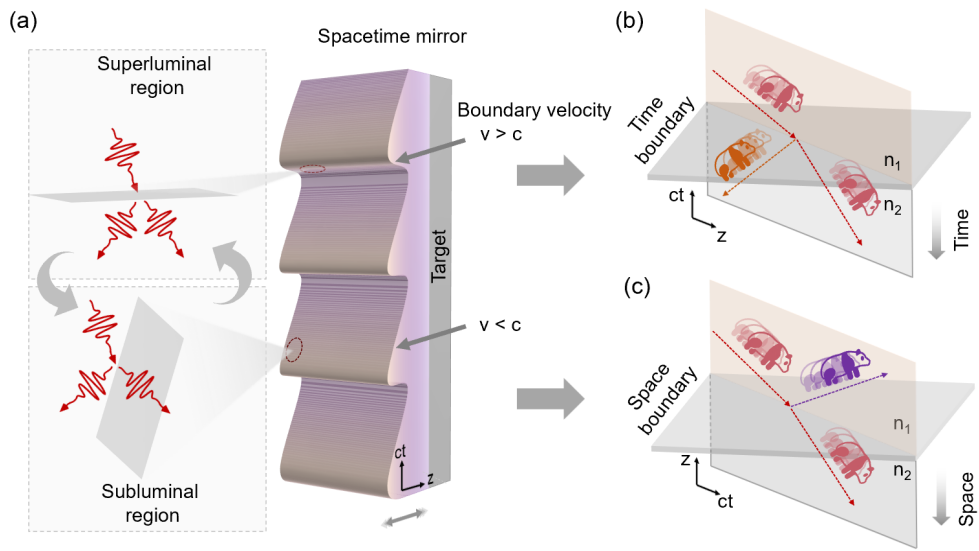

**Figure 1**: Construction of a spacetime mirror through the laser-induced relativistically oscillating plasma sheets. (a) The variations in speed of the spacetime mirror correspond to subluminal (space reflection) and superluminal (time reflection) reflection of the incident field. (b) Time reflection and refraction indicate that the wave vector of the reflected field aligns with that of the incident and refracted field, thereby conserving momentum. The main feature of time reflection is the generation of new photons from the time boundary, and energy conservation is violated. (c) In contrast, reflection and refraction in space reveals that the wave vector of the reflected field is directed opposite to that of the incident and refracted lights, with the optical frequency being conserved.

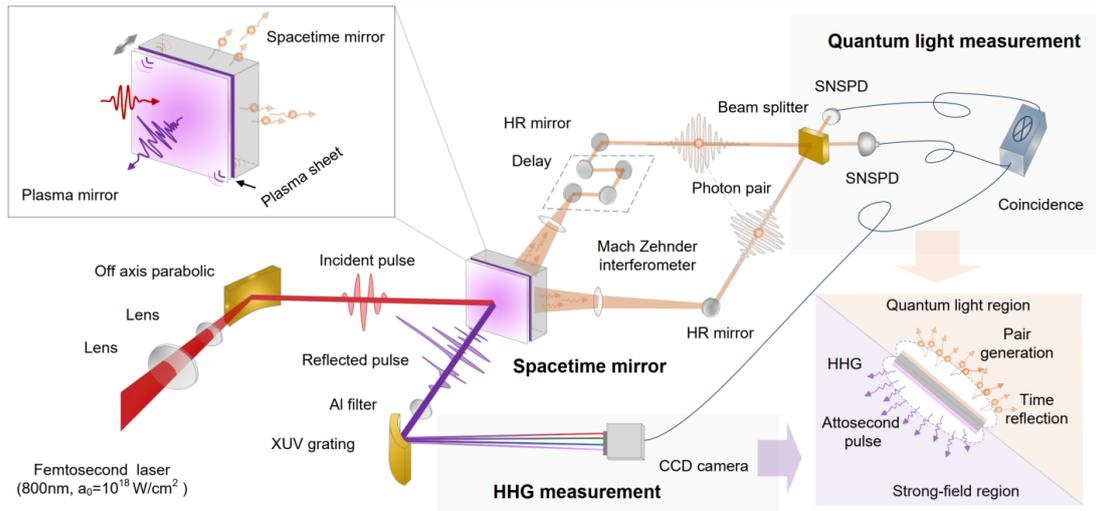

**Figure 2:** Schematic setup for combined quantum light measurement and strong field high-harmonic generation (HHG) measurement. A relativistic laser interacts with an $Sio_2$ target coated with a metal thin film, which excites relativistic oscillating mirror (ROM) behavior as a spacetime mirror. HHG is detected in the strong field regimes, while on the backside of the target paired photons generated by the spacetime mirror are detected using HOM interference setup. The reflected HHG spectrum and photon pairs are altogether imaged by a coincidence measurement. (Left panel) The relativistic laser fully ionizes the target surface and facilitates the formation of a dense plasma to creates the superluminal spacetime mirror. (Right panel) This spacetime mirror can generate HHG in in the strong-field region and squeeze the dielectric vacuum electromagnetic field to produce entangled photons in the quantum light region .

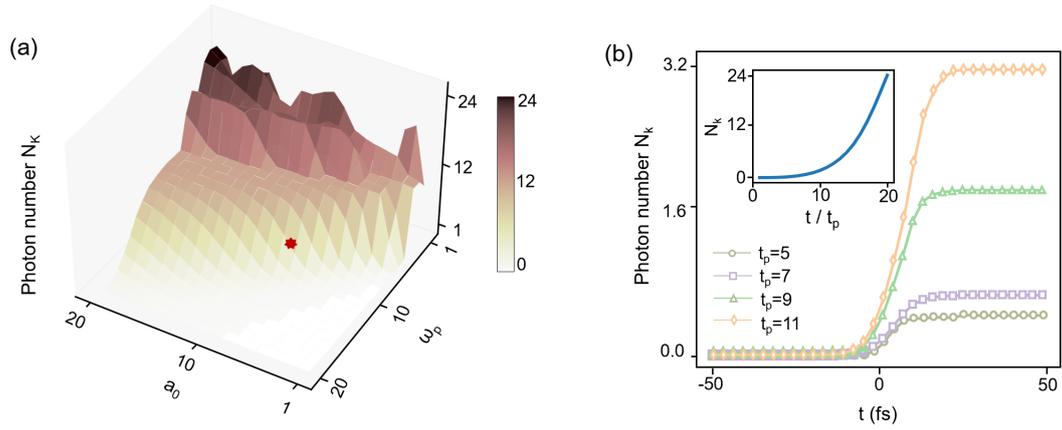

**Figure 3**:Characteristics of quantum light generation in terms of physical parameters such as laser intensity, plasma density, and interaction time (laser pulse width). (a) Higher laser intensity creates more photon pairs, while lower plasma density, the more photons can also enhances pair generation. The laser pulse width is chosen at $10T_L$. (b) longer interaction time results in greater photon production, with the photon number increasing exponentially as the pulse width increases (inset). Specifically, as the pulse width increases from $4T_L$ to $12T_L$, photon count rises from 0.1 to 4.

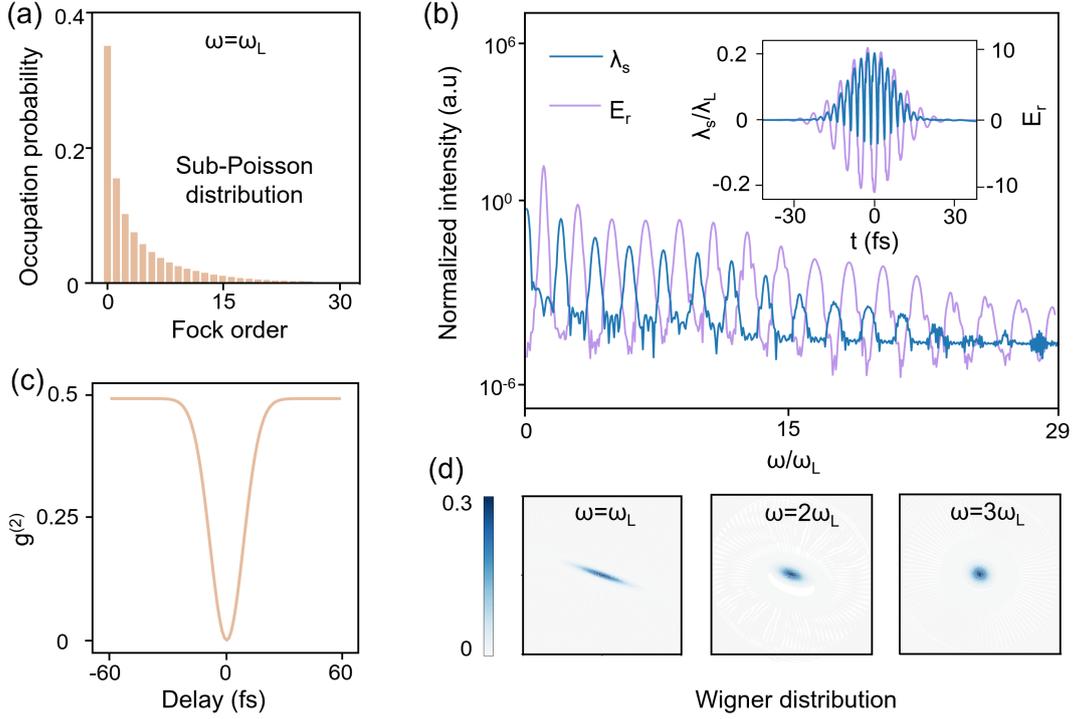

**Figure 4**: The common origin of quantum light generation and high harmonics generation from spacetime mirrors. (a) The Fock state distribution of quantum light with frequency $\omega = \omega_L$ shows a sub-Poisson distribution. (b) The orders of harmonic generation is odd, while the plasma oscillation is even. This distinction arises because HHG combines the plasma sheet oscillation frequency with the driving frequency. (c) shows the characteristics of the second-order correction function $g^{(2)}$ recorded by HOM interference. When there is no delay between two arms, the dip feature of $g^{(2)}$ indicates that the two emitted photons are entangled. Notably, these photon pairs at high harmonic orders are also entangled. (d) shows the characteristics of squeezed vacuum state of the 1st-3rd order harmonics in the representation of Wigner functions. The squeezing effect is weaker for a higher order.

# Supplementary Material:

# Superluminal spacetime boundary, time reflection and quantum light generation from relativistic plasma mirrors

Chenhao Pan[1,2], Xinbing Song[3], Yang Cao[4], Li Xiong[5], Xiaofei Lan[5], Shaoyi Wang[6*], Yuxin Leng[1,2*], Yiming Pan[1*]

[1]*School of Physical Science and Technology and Center for Transformative Science, ShanghaiTech University, Shanghai 201210, China*
[2]*State Key Laboratory of High Field Laser Physics and Chinese Academy of Sciences Center for Excellence in Ultra-Intense Laser Science, Shanghai Institute of Optics and Fine Mechanics, Chinese Academy of Sciences, Shanghai 201800, China*
[3]*School of Physics, Beijing Institute of Technology, Beijing 100081, China*
[4]*Physics Department, Technion, Haifa, 320000, Israel*
[5]*School of Physics and Astronomy, China West Normal University, Nanchong 637002, China*
[6]*Science and Technology on Plasma Physics Laboratory, Research Center of Laser Fusion, China Academy of Engineering Physics, Mianyang 621900, China*

Including:

Section 1: Spacetime mirror based on the relativistic oscillating mirror

Section 2: Spontaneous pair generation for a cavity with oscillating boundary

Section 3: The Hong-Ou-Mandel measurement setup

# Section 1: Spacetime mirror based on the relativistic oscillating mirror

In this section, the model of spacetime mirror based on the relativistic oscillating mirror (ROM) is developed. This basic idea was first proposed by Bulanov *et al*. Here, we derive it in detail.

We assume that the electron distribution of the plasma target has a sharp boundary described by the moving coordinate $\lambda_s(x,t)$. Then electron density is then expressed as:

$$n(x,t) = \Theta(x - \lambda_s(x,t)) \tag{1}$$

Where $\Theta(x)$ is the step function. The configuration is depicted in Fig. S1. When the laser field irritates the plasma, the motion of the particle with mass $m$ and charge $q$ is governed by the relativistic Lorentz equation:

$$\frac{d}{dt}\boldsymbol{p} = q(\boldsymbol{E} + \boldsymbol{v} \times \boldsymbol{B}) \tag{2}$$

where $\boldsymbol{E} = -(1/c)\partial_t \boldsymbol{A}(x,t) - \nabla\Phi(x,t)$ and $\boldsymbol{B} = \nabla \times \boldsymbol{A}(x,t)$ is the electric (magnetic) field of the driving laser with $\boldsymbol{A}(x,t)$ is the laser vector potential and $\Phi(x,t)$ is the static electric potential along $\boldsymbol{x}$ axis, which has been defined in the main text, the velocity of the particle is denoted by $\boldsymbol{v}$, and $\boldsymbol{p} = m\gamma\boldsymbol{v}$ is its relativistic momentum. The Lorentz factor is given by $\gamma = [1 - (v/c)^2]^{-1/2} = [1 + (p^2/m^2c^2)]^{1/2}$ is the Lorentz factor, the total time derivative of $\boldsymbol{p}$ is given as $d_t\boldsymbol{p} = \partial_t\boldsymbol{p} + (\boldsymbol{v}\cdot\nabla)\boldsymbol{p}$. Thus, the kinetic energy balance is:

$$\frac{d}{dt}(mc^2\gamma) = q\big(-v_x\partial_x\Phi(x,t) - \boldsymbol{v}_\perp \cdot \partial_t \boldsymbol{A}(x,t)\big) \tag{3}$$

where $v_x\partial_x\Phi(x,t)$ claims the longitudinal energy change, while $\boldsymbol{v}_\perp \cdot \partial_t \boldsymbol{A}(x,t)$ depicts the transverse. Additionally, the continuity equation is:

$$\partial_t \rho + \nabla \cdot \boldsymbol{j} = \partial_t\rho + \partial_x j_l = 0 \tag{4}$$

with $\rho$ is the particle density and $\boldsymbol{j} = -e(n_e\boldsymbol{v}_e - Zn_i\boldsymbol{v}_i)$, where $e$ is the unit charge and the initial density distribution of electrons and ions is given by $n_e(x) = Zn_i(x) = n_0\Theta(x)$. The longitudinal current is defined as $j_l = -e(n_e v_{x,e} - Zn_i v_{x,i})$.

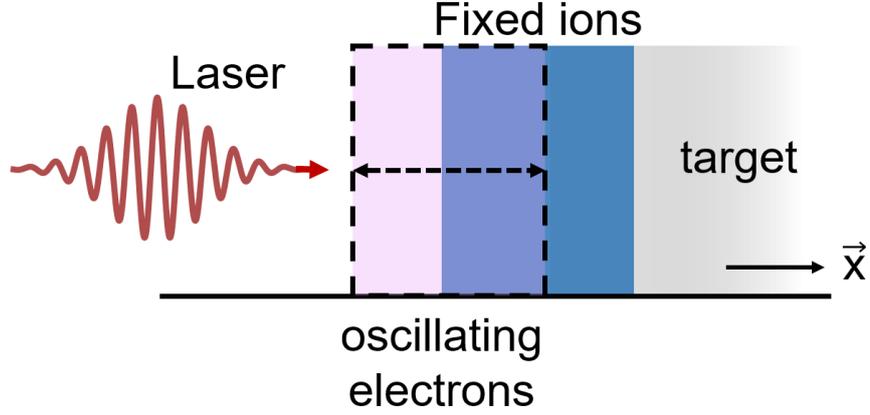

**Fig. S1**: Schematic plot of the spacetime mirror model, the electron density (dash line) of the vacuum-plasma surface at $\lambda_s(x,t)$, oscillating relatively to the fixed ions density. This process is localized at the front surface.

We notice that, this system needs to be analyzed in both longitudinal and transverse directions, thus, we separate the relativistic Lorentz equation into components along each directions, for the longitudinal direction, we have:

$$\frac{d}{dt}p_l = q\left(-\frac{\partial}{\partial x}\Phi(x,t) + v_\perp \cdot \partial_x A(x,t)\right) \quad (5)$$

For the transverse direction, we have:

$$\frac{d}{dt}p_\perp = -q \cdot \partial_t A(x,t) \quad (6)$$

This indicates that, the particles motion on the transverse direction is totally controlled by the electric field of the laser. Thus, we can estimate the electron and ion velocities if we assume the electric field $E(x,t) = E_0 \sin(k_L x - \omega_L t)$. The corresponding velocities of electrons and ions are given by:

$$\begin{cases} v_{\perp,e} = -\dfrac{eE_0}{m_e \omega_L} \\ v_{\perp,i} = \dfrac{ZeE_0}{m_i \omega_L} \end{cases} \quad (7)$$

It is known that $Z/m_i \ll 1/m_e$, which implies that the ions are approximately unmoved. For a laser working at $800nm$ to drive the electrons transverse speed $v_{\perp,e} \sim$ $c$, we can evaluate that $E_0 \sim 10^{12} V/m$. The normalized relativistic intensity $a_0 = eE_0/(m\omega_L c) \sim 1$, we can have the intensity density $I\lambda_L^2 \sim 10^{18} W/cm^2$.

In this situation, we need rewrite the Lorentz factor $\gamma = [1 - (v_x/c) - (v_\perp/c)]^{-1/2}$, with $v_x$ is varying with time. So, for the longitudinal direction:

$$\frac{d}{dt}p_l = \left(\frac{d}{dt}\gamma\right)m\left(\frac{d}{dt}\lambda_s(t)\right) + \frac{d}{dt}\left(\frac{d}{dt}\lambda_s(t)\right)m\gamma \tag{8}$$

With $d_t\gamma = -n_e e(-v_x \partial_x \Phi(x,t) - v_\perp \cdot \partial_t A(x,t))/m_e c^2$. Finally, we can obtain the plasma surface moving equation:

$$\frac{d}{dt}(\dot{\lambda}_s(t)\gamma) = -n_0\lambda_s(t)\Theta(\lambda_x(t)) + \frac{a}{\gamma}\frac{\partial}{\partial x}a \tag{9}$$

Here, $\lambda_s(t)$ is the displacement of the plasma surface with the ions motion are ignored. Then we can approximately obtain that $\lambda_s(t) = \lambda_s \sin(2\omega_L t)$, with the pondermotive force driving. Given that $d_t\lambda_s(t) \approx 2\omega_L\lambda_s \sim c$, we find $\lambda_s \sim c/2\omega_L \sim 1/10\lambda_L$.

Our concern here is to investigate the spacetime characteristic of the vacuum-plasma surface. The particle-in-cell (PIC) simulations, obtained with 1D code (EPOCH) [61], which described the variation of vacuum-plasma surface with respect to both space and time. We discuss a series of cases, varying laser field strength and plasma density. For instance, we start with the case, whose parameters are used to discuss the quantum light generation. The laser beam with $a_0 = 10$, is normally incident on the plasma target. Electron density is visualized using the colormap (as shown in Fig. S2), while the ions density is not displayed. The sharp interface is fixed at $x = 0$. One can see that the electron density oscillates in both space and time. However, the density decreases in the oscillating area, thus, we depict the contour lines for the electron density above the critical density ($\omega_p/\omega_L > 1$). These contour lines represent the oscillating spacetime mirror. We can notice that contour lines oscillation contains multi frequencies. Some areas may show the characteristic of superluminal speed as compared to the speed of light line (red dash line). Also, our numerical method follows [62] meets well with the PIC simulation result

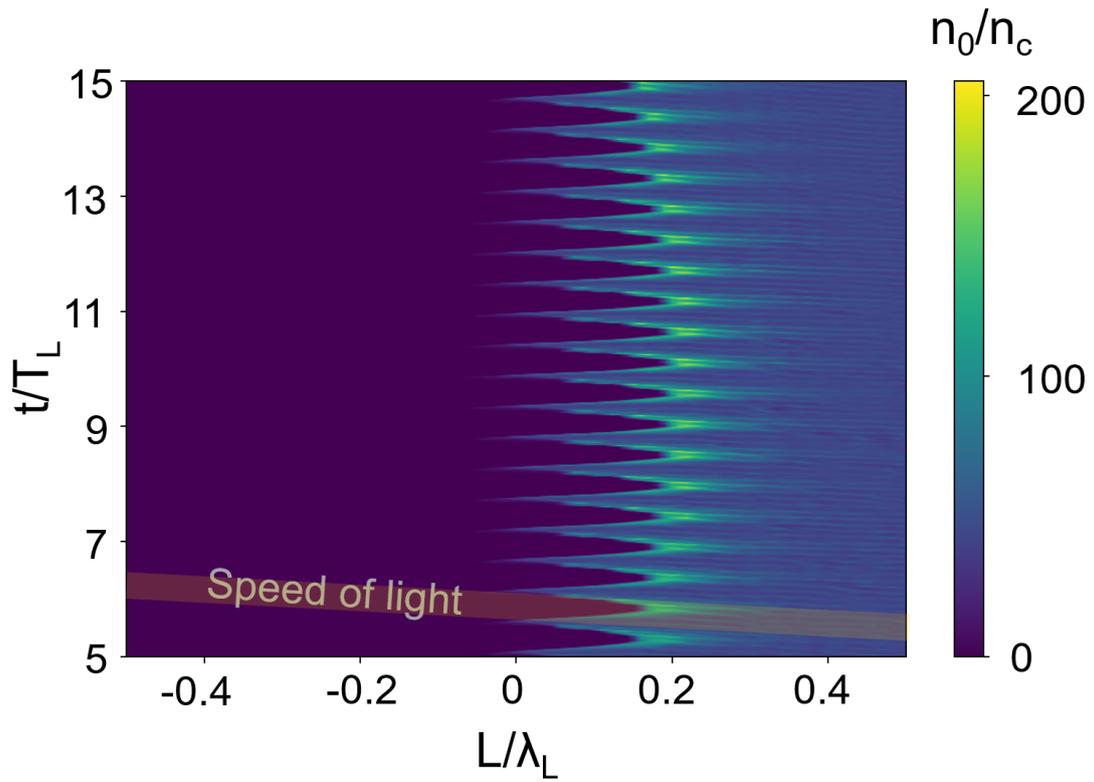

**Fig. S2**: Particle-in-cell (PIC) result of the plasma spacetime mirror with $a_0 = 10$, $n_0 = 100 n_c$ and $t_p = 10 T_L$. The red line shows the speed of light, which can compare it to the spacetime mirror and see that spacetime mirror can achieve the effect of superluminal speed.

# Section 2: Spontaneous pair generation for a cavity with oscillating boundary

We consider a one-dimensional cavity formed by two perfectly reflecting mirrors (as shown in Fig. S3). One of the mirrors is fixed at position $L_0$ and consists of the high reflection film. The other is the oscillating one formed by the plasma spacetime mirror. With the Coulomb gauge, the vector potential $A(x,t)$ within the cavity satisfies the wave equation obtained from Maxwell equations, where $c = 1$ is the speed of light in the dielectric target:

$$\left(\nabla^2 - \frac{1}{c^2}\frac{\partial^2}{\partial t^2}\right) A(x,t) = 0 \tag{1}$$

We can decompose the vector potential into two complex components: $A(x,t) = A^{(+)}(x,t) + A^{(-)}(x,t)$ with $A^{(-)} = \left(A^{(+)}\right)^*$. Once we set both of the mirrors are stable, we can describe the field restricted to a certain volume of space and expand the vector potential in terms of a set of orthogonal mode functions:

$$A^{(+)}(x,t) = \sum_n C_n u_n(x) e^{-i\omega_n t} \tag{2}$$

And $u_k(x)$ and $\omega_k$ satisfy the Helmholtz equation:

$$(\nabla^2 - k_n^2) u_n(x) = 0 \tag{3}$$

Subject to the boundary condition $u_n(0) = u_n(L_0) = 0$, and $k_n = 2\pi k/L_0 = \omega_n/c$ is the wave vector and $\omega_n$ is the wave frequency. Additionally, $u_n(x)$ satisfies the transversality condition $\nabla \cdot u_n(x) = 0$, and the complete orthonormal condition:

$$\int_0^{L_0} u_n^*(x) u_{n'}(x) dx = \delta_{nn'} \tag{4}$$

Following the formulism of the second quantization procedure, the field operator can be constructed:

$$\hat{A}(x,t) = \sum_n \sqrt{\frac{\hbar}{2\omega_n \epsilon}} \left[\alpha_n u(x) e^{-i\omega_n t} + \alpha_n^+ u^*(x) e^{i\omega_n t}\right], \tag{5}$$

where $\alpha_k$ and $\alpha_k^+$ is the annihilation and creation operators satisfy the boson commutation relations:

$$[\alpha_n, \alpha_{n'}^+] = i\hbar\delta_{nn'} \quad [\alpha_n, \alpha_{n'}] = [\alpha_n, \alpha_{n'}^+] = 0 \tag{6}$$

When the boundary begins to oscillate, we impose the periodic boundary condition:

$$A(0,t) = A(L(t),t) \tag{7}$$

We first define the "instantaneous" set of mode basis $\phi_k(x,t)$ [63]:

$$(\nabla^2 - k_n^2(t))\phi_n(x,t) = 0 \tag{8}$$

Subject to the boundary condition:

$$\phi_n(0,t) = \phi_n(L(t),t) = 0 \tag{9}$$

where $k_n(t) = 2\pi n/L(t)$. This definition helps us to construct the new set of bases can be considered unchanged at each instantaneous moment. Such bases are orthonormal:

$$\int_0^{L(t)} \phi_n(x,t)\phi_m(x,t)dx = \delta_{nm} \tag{10}$$

and are complete, from which we can have $\phi_n(x,t) = \sqrt{2/L(t)}\sin(ik_n(t)x)$. Hence, the field operator $\hat{A}(x,t)$ can be expanded in terms of the instantaneous basis at any instant $t$:

$$\hat{A}(x,t) = \sum_n \hat{Q}_n(t)\phi_n(x,t) \tag{11}$$

The Lagrangian density of the system is given by:

$$\ell(x,t) = \frac{1}{2}\left(|\hat{E}(x,t)|^2 - |\hat{B}(x,t)|^2\right) \tag{11}$$

with $\hat{E}(x,t) = (1/c)\partial_t \hat{A}(x,t)$ and $\hat{B}(x,t) = \nabla \times \hat{A}(x,t)$, and we can have the Lagrangian of the system is:

$$\mathcal{L}(x,t) = \int_0^{L(t)} \frac{1}{2}\sum_k \left\{\dot{\hat{Q}}_n^2(t)\phi_n^2(x,t) + \hat{Q}_n^2(t)\dot{\phi}_n^2(x,t) + \dot{\hat{Q}}_n(t)\hat{Q}_n(t)[\phi_n(x,t)\dot{\phi}_n(x,t)]\right.$$
$$\left. + \hat{Q}_n(t)\dot{\hat{Q}}_n(t)[\phi_n(x,t)\dot{\phi}_n(x,t)] - c^2\hat{Q}_n(t)[\nabla\phi_n(x,t)]^2\right\}dx \tag{12}$$

We define $\dot{\hat{Q}}_n(t) = \partial_t Q_n(t)$, $\dot{\phi}_n(x,t) = \partial_t \phi_n(x,t)$, with the orthogonal and complete condition:

$$\mathcal{L}(x,t) = \frac{1}{2}\sum_n \left\{ \dot{\hat{Q}}_n^2(t) + \dot{\hat{Q}}_n(t)\hat{Q}_n(t)\left[\int_0^{L(t)} \phi_n(x,t)\dot{\phi}_n(x,t)\,dx\right] \right.$$

$$+ \hat{Q}_n(t)\dot{\hat{Q}}_n(t)\left[\int_0^{L(t)} \phi_n(x,t)\dot{\phi}_n(x,t)\,dx\right]$$

$$\left. - \hat{Q}_n(t)\left[\int_0^{L(t)} c^2[\nabla\phi_n(x,t)]^2 - [\dot{\phi}_n(x,t)]^2\,dx\right] \right\} \quad (13)$$

Introducing the canonical conjugate momentum $\hat{P}_n(t) = \partial \ell(x,t)/\partial \dot{Q}_n(t)$, we can obtain the system Hamiltonian with the Legendre transformation $H_{EM} = \int_0^{L(t)} \dot{\hat{Q}}_n(t)\frac{\partial}{\partial \dot{Q}_n(t)}\ell(x,t)dx - \mathcal{L}(x,t)$:

$$H_{EM} = \frac{1}{2}\sum_n \left\{ \hat{P}_n^2(t) + \hat{P}_n(t)\hat{Q}_n(t)\left[\int_0^{L(t)} \phi_n(x,t)\dot{\phi}_n(x,t)\,dx\right] \right.$$

$$+ \hat{Q}_n(t)\hat{P}_n(t)\left[\int_0^{L(t)} \phi_n(x,t)\dot{\phi}_n(x,t)\,dx\right]$$

$$\left. + \hat{Q}_n^2(t)\left[\int_0^{L(t)} [\nabla\phi_n(x,t)]^2\,dx\right] \right\} \quad (14)$$

where the coupling terms are defined as:

$$G_{nn}(t) = \int_0^{L(t)} \phi_n(x,t)\dot{\phi}_n(x,t)\,dx$$

$$G_{nn}^2(t) = \int_0^{L(t)} \dot{\phi}^2{}_n(x,t)\,dx$$

$$G_{mn}(t) = \int_0^{L(t)} \phi_n(x,t)\dot{\phi}_m(x,t)\,dx, (m \neq n)$$

with the property $G_{mn}(t) = -G_{nm}(t)$, then the Hamiltonian $H_{EM}$ can be simplified as:

$$H_{EM} = \frac{1}{2}\sum_n \left\{\hat{P}_n^2(t) + [\hat{P}_n(t)\hat{Q}_n(t) + \hat{Q}_n(t)\hat{P}_n(t)] \cdot G_{nn}(t) \right.$$
$$\left. + \hat{Q}_n^2(t)\left[\int_0^{L(t)} [\nabla\phi_n(x,t)]^2 \, dx\right]\right\} + \sum_{\substack{m,n \\ m \neq n}} [\hat{P}_m(t)\hat{Q}_n(t)] \cdot G_{mn}(t) \quad (15)$$

Further, we give the detailed form of $\int_0^{L(t)}[\nabla\phi_n(x,t)]^2 \, dx$ with:

$$\nabla\phi_n(x,t) = k_n(t)\cos(k_n(t)x) \quad (16)$$

And

$$\dot{\phi}_n(x,t) = -\frac{\sqrt{2}}{2}\frac{\dot{L}(t)}{L^{\frac{3}{2}}(t)}\sin(k_n(t)x) + \sqrt{\frac{2}{L(t)}}\frac{\dot{L}(t)}{L(t)}k_n(t)x\cos(k_k(t)x) \quad (17)$$

from which we can have the final Hamiltonian:

$$H_{EM} = \frac{1}{2}\sum_n \{\hat{P}_n^2(t) + \hat{Q}_n^2(t)[\omega_n^2(t)] + [\hat{P}_n(t)\hat{Q}_n(t) + \hat{Q}_n(t)\hat{P}_n(t)] \cdot G_{nn}(t)\}$$
$$+ \sum_{\substack{m,n \\ m \neq n}} [\hat{P}_n(t)\hat{Q}_n(t)] \cdot G_{kj}(t)$$

We now introduce the "instantaneous" creation and annihilation operators:

$$\hat{a}_n = \frac{1}{\sqrt{2\hbar\omega_n(t)}}[\omega_n(t)\hat{Q}_n(t) + i\hat{P}_n(t)]$$

$$\hat{a}_n^+ = \frac{1}{\sqrt{2\hbar\omega_n(t)}}[\omega_n(t)\hat{Q}_n(t) - i\hat{P}_n(t)]$$

with the commutation relation:

$$[\hat{Q}_n(t), \hat{P}_{n'}(t)] = i\hbar\delta_{nn'}, [\hat{Q}_n(t), \hat{Q}_m(t)] = [\hat{P}_n(t), \hat{P}_m(t)] = 0$$

and

$$\hat{Q}_n(t) = \sqrt{\frac{\hbar}{2\omega_n(t)}}(\hat{a}_n + \hat{a}_n^+)$$

$$\hat{P}_n(t) = -i\sqrt{\frac{\hbar\omega_n(t)}{2}}(\hat{a}_n - \hat{a}_n^+)$$

We then can obtain the final form of the system Hamiltonian:

$$H_{EM} = \hbar \sum_n \{\omega_n(t)a_n^+ a_n + i\xi_n(t)[(a_n^+)^2 - a_n^2]\}$$

$$+ \frac{i\hbar}{2} \sum_{\substack{m,n \\ m \neq n}} \mu_{mn}(t)(a_n^+ a_m^+ - a_n a_m + a_n^+ a_m - a_n^+ a_m)$$

where $\mu_{mn}(t) = \left(\frac{\omega_m(t)}{\omega_n(t)}\right)^{\frac{1}{2}} (-1)^{(m+n)} \frac{mn}{m^2-n^2} \frac{\dot{L}(t)}{L(t)}$, $\xi_n(t) = \dot{L}(t)/4L(t)$.

This Hamiltonian describes three distinct physical processes in our system. The first is the zero-photon process, characterized by $a_n^+ a_m$ terms, which describe photons are scattered from one mode to another without changing total photon number. The second and the third are both two photon processes, characterized by terms $(a_n^+)^2$ and $a_n^+ a_m^+$, so that photon pairs can be created from the vacuum state. It should be noted that the contribution of these processes are governed by the time-dependent coefficients $\xi_k(t)$ and $\mu_{kj}(t)$.

To clearly identify which physical process dominates in this system, we numerically compare the magnitudes of $|\xi_k(t)|$ for single mode squeezing and $|\mu_{kj}(t)|$ for both two-mode squeezing and two-mode scattering, as shown in Fig. S4. Each order of $L(t)$ and $\dot{L}(t)$ is extracted from the numerical simulation results of the spacetime mirror. From these results, we observe that at lower orders, the coefficient for single-mode squeezing is approximately 1000 times greater than that for the two-mode process, indicating that the system is dominated by the single-mode squeezing process.

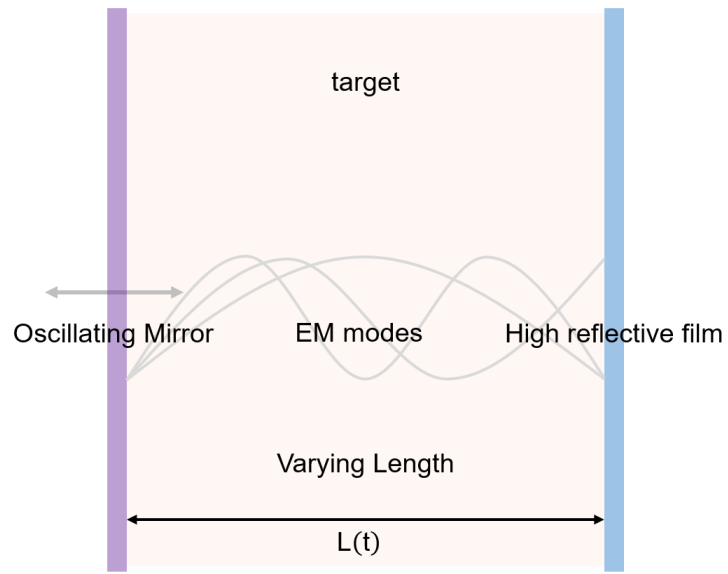

**Fig. S3:** Schematic plot of the spacetime mirror model forming the oscillating cavity, the oscillating mirror is conformed with plasma, the length of the cavity $L(t)$ varying with spacetime mirror trajectory.

## Section 3: The Hong-Ou-Mandel measurement setup

Two-photon interference, commonly referred to as the Hong-Ou-Mandel (HOM) effect, is considered the "heart of quantum mechanics" due to its quantum nature, with absolutely no analogue in classical physics [64]. The HOM effect enables distinguish between entangled photons and classical light clearly, allowing us verify the photons received by the detectors are generated by squeezing vacuum rather than the leakage of the incident light.

The key element of the HOM effect is the action of the beam splitter (BS), providing the mixing between the two input modes. As a starting point of the HOM effect, we introduce the description of operational action of the BS. A BS is an optical device with two input ports, labelled $a$ and $b$, and two output ports, labelled $c$ and $d$, as shown in Fig. S5(b). Beams incident on a BS at the point $a$ or $b$ is split between output ports $c$ and $d$ in proportions depending on reflectance $r$ and transmittance $t$. Here, we considered the balanced BS where $|r| = |t| = 1/\sqrt{2}$, also known as a 50:50 BS [64].

We now describe the quantum behavior of the BS using the second quantization formalism. This is done by employing four sets of bosonic annihilation and creation operators $(\hat{l}, \hat{l}^+)$ ($l = a, b, c, d$) to represent electromagnetic fields in mode $i$, and must satisfy the standard bosonic commutation relation: $[\hat{l}, \hat{l'}^+] = \delta_{ll'}$, while $\delta_{ll'}$ is the Kronecker delta symbol. The operation of a 50:50 BS can then be described in terms of field operators by:

$$\begin{cases} \hat{a} = \frac{1}{\sqrt{2}}(\hat{c} + \hat{d}) \\ \hat{b} = \frac{1}{\sqrt{2}}(\hat{c} - \hat{d}) \end{cases} \quad (1)$$

The combined two-photon state before arriving at the BS, as the input state is:

$$|\psi_{in}\rangle_{ab} = \hat{a}_j^+ \hat{b}_k^+ |0\rangle \quad (2)$$

After the action of BS operator, the output state becomes:

$$|\psi_{out}\rangle_{cd} = \hat{U}_{BS}(\hat{a}_j^+ \hat{b}_k^+|0\rangle) = \frac{1}{\sqrt{2}}(\hat{c}_j^+ + \hat{d}_j^+)\frac{1}{\sqrt{2}}(\hat{c}_k - \hat{d}_k)|0\rangle$$

$$= \frac{1}{2}(\hat{c}_j^+ \hat{c}_k^+ + \hat{c}_k^+ \hat{d}_j^+ - \hat{c}_j^+ \hat{d}_k^+ - \hat{d}_k^+ \hat{d}_j^+)|0\rangle$$

From the analysis above, the effective Hamiltonian can create photon pairs with same frequency spectral, polarization, temporal duration, and transverse spatial mode. Due to the phase matching condition, we can collect the photons at the symmetry direction as shown in Figure 2. In this scenario, the output state is given by:

$$|\psi_{out}\rangle = (\hat{c}_k^+ \hat{c}_k^+ + \hat{c}_k^+ \hat{d}_k^+ - \hat{c}_k^+ \hat{d}_k^+ - \hat{d}_k^+ \hat{d}_k^+)|0\rangle = (c_k^+ \hat{c}_k^+ - \hat{d}_k^+ \hat{d}_k^+)|0\rangle$$

$$= \frac{1}{\sqrt{2}}(|2\rangle_a - |2\rangle_b)$$

Here, the coincidence probability of detecting one photon in each output mode is $p = 0$. The result indicates that when two indistinguishable photons interfere at a $50{:}50$ BS, the amplitude for "both transmitted" and "both reflected" perfectly cancel out.

For further analysis, the spectral profile of the photons takes into count, since the photons generation only occurs within the interaction period, thus, the photons with second order oscillation have the same spectral amplitude function as driving laser field.

$$f_i(\omega) = \frac{1}{\pi^{\frac{1}{4}}\sqrt{\sigma_i}} e^{-\frac{(\omega-\omega_i)^2}{2\sigma_i^2}}, (i = a, b) \tag{4}$$

Where $\omega_i$ is the center frequency of photon $i$, $\sigma_i$ defines the spectral width, and the normalization was chosen $\int d\omega |f_i(\omega)|^2 = 1$. By controlling the time delays between two such photons, it is possible to tune their level of distinguishability. This is shown schematically in Fig. S5 (a).

We are interested in how the coincidence probability changes as a function of the overlap between the photons. We thus introduce a time delay in line $b$. In practice, this might be done by sending the photon in line $b$ through an adjustable delay line that introduces a phase shift:

$$\hat{b}^+(\omega) \to \hat{b}^+(\omega) e^{-i\omega\tau} \tag{5}$$

the time-delayed state is then:

$$|\psi_{td}\rangle_{ab} = \int d\omega f_a(\omega) \hat{a}^+(\omega) \int d\omega f_b(\omega) \hat{b}^+(\omega) e^{-i\omega\tau} |0\rangle \tag{6}$$

After passing through the BS, the output state is:

$$|\psi_{out}\rangle_{cd} = \hat{U}_{BS} |\psi_{td}\rangle_{ab}$$

$$= \frac{1}{2}\int d\omega f_a(\omega)\left(\hat{c}^+(\omega)+\hat{d}^+(\omega)\right)\int d\omega f_b(\omega)\left(\hat{c}^+(\omega)-\hat{d}^+(\omega)\right)e^{-i\omega\tau}|0\rangle$$

$$= \frac{1}{2}\int d\omega f_a(\omega)\int d\omega f_b(\omega)e^{-i\omega\tau}$$

$$\times \left(\hat{c}^+(\omega)\hat{c}^+(\omega)+\hat{c}^+(\omega)\hat{d}^+(\omega)-\hat{c}^+(\omega)\hat{d}^+(\omega)+\hat{d}^+(\omega)\hat{d}^+(\omega)\right)$$

The coincidence probability of detecting one photon in each mode is:

$$p = Tr[|\psi_{out}\rangle\langle\psi_{out}|\hat{P}_c\otimes\hat{P}_d] = \langle\psi_{out}|\hat{P}_c\otimes\hat{P}_d|\psi_{out}\rangle \quad (7)$$

With $\begin{cases}\hat{P}_c = \int d\omega \hat{c}^+(\omega)|0\rangle\langle 0|\hat{c}(\omega) \\ \hat{P}_d = \int d\omega \hat{d}^+(\omega)|0\rangle\langle 0|\hat{d}(\omega)\end{cases}$, and with the profile defined above, the probability reads:

$$p = \frac{1}{2} - \frac{1}{2\pi\sigma^2}\left(\int d\omega e^{-\frac{(\omega-\omega_a)^2}{2\sigma_a^2}}e^{-\frac{(\omega-\omega_b)^2}{2\sigma_b^2}}e^{-i\omega\tau}\right)\left(\int d\omega e^{-\frac{(\omega-\omega_a)^2}{2\sigma_a^2}}e^{-\frac{(\omega-\omega_b)^2}{2\sigma_b^2}}e^{-i\omega\tau}\right) (8)$$

with the Fourier transformation, we can finally obtain the coincidence probability:

$$p = \frac{1}{2} - \frac{1}{2}e^{-\sigma^2\tau^2} \quad (9)$$

This result is shown in Figure 4 (c), meets well with the numerical simulation.

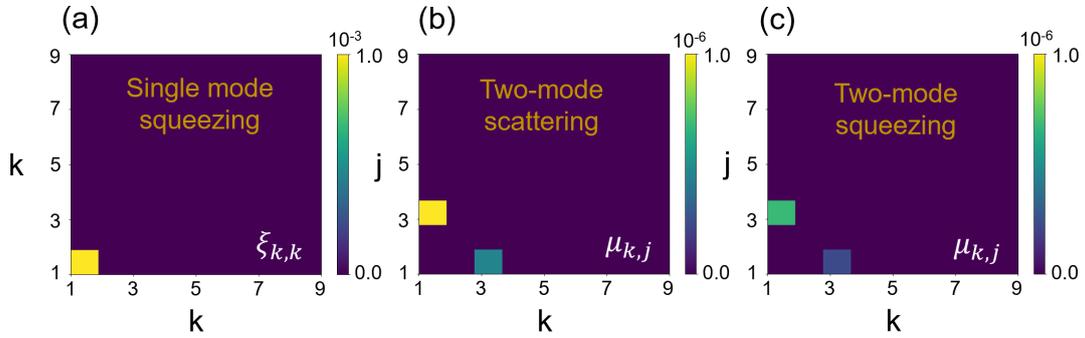

**Fig. S4**: The numerical values of the coefficients for (a) single-mode squeezing, (b) two-mode scattering and (c) two-mode squeezing. We compare the three coefficients and find that the single mode squeezing would dominate.

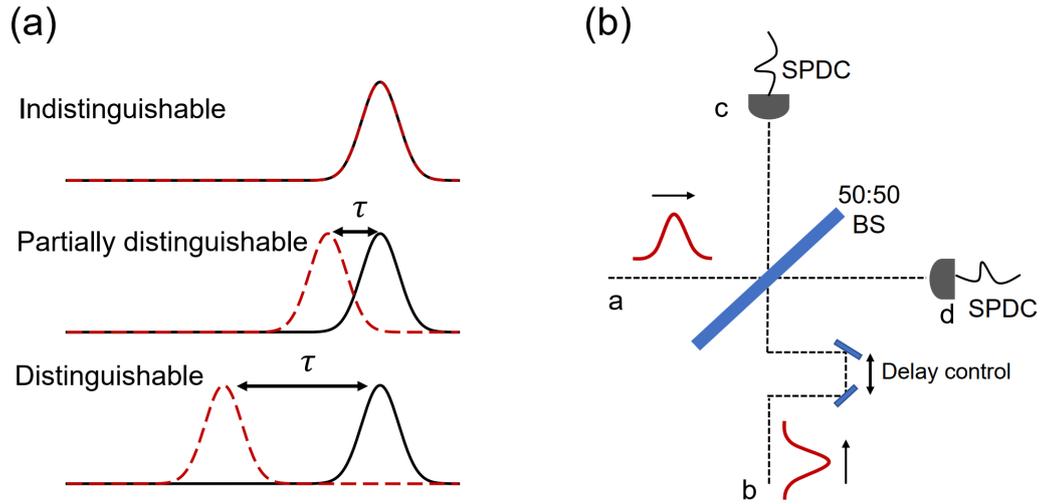

**Fig. S5:** Schematic plot of Hong-Ou-Mandel interference. (a) Adjusting the distinguishability by controlling the delay $\tau$. (b) the setup of the HOM interference, the delay control is realized by two relatively moving mirrors, single photon detectors (SPDC) counting the photons after two photon pulses hitting on the beam splitter (BS).